\documentstyle[bbm,amsfonts,epsfig,12pt,here]{article}

\newcommand {\eqref} [1] {(\ref {#1})}

\newcommand {\slsh} [1] {\not{\hbox{\kern-2pt${#1}$}}}

\def\drawbox#1#2{\hrule height#2pt
         \hbox{\vrule width#2pt height#1pt \kern#1pt
               \vrule width#2pt}
               \hrule height#2pt}

\def\Asym#1#2{\vcenter{\vbox{\drawbox{#1}{#2}
               \kern-#2pt       % line up boxes
               \drawbox{#1}{#2}}}}

\newcommand{\nunez}{N$\mathrm{\acute{u}}\tilde{\mathrm{n}}$ez }

\newcommand {\beq} {\begin{equation}}
\newcommand {\eeq} {\end{equation}}
  \newcommand {\ber}{\begin{eqnarray*}}
  \newcommand {\eer} {\end{eqnarray*}}
\newcommand {\bea}{\begin{eqnarray}}
  \newcommand {\eea} {\end{eqnarray}}

\newcommand{\Dslash}{\,{\raise.15ex\hbox{/}\mkern-12mu D}}

\begin{document}

%%%%%%%%%%%%%%%%%%%%%%%%%%%%%%%%%%%%%%%%%%%%%%%%%%%%%%%%%%%%%%%%%%%%%%%%%%%%%%%

\begin{titlepage}

\begin{center}
\vspace{1in}
\large{\bf Beyond The Quenched (or Probe Brane) Approximation in Lattice (or Holographic) QCD}\\
\vspace{0.4in}
\large{Adi Armoni}\\
\small{\texttt{a.armoni@swan.ac.uk}}\\
\vspace{0.2in}
\large{\emph{Department of Physics, Swansea University}\\ 
\emph{Singleton Park, Swansea, SA2 8PP, UK.}\\}
\vspace{0.5in}
\end{center}

\abstract{We propose a method to improve the quenched approximation. The method, based on the worldline formalism, takes into account effects of quark loops. The idea is mostly useful for AdS/CFT (holographic) calculations. To demonstrate the method we estimate screening (string breaking) effects by a simple holographic calculation. }

\end{titlepage}

%%%%%%%%%%%%%%%%%%%%%%%%%%%%%%%%%%%%%%%%%%%%%%%%%%%%%%%%%%%%%%%%%%%%%%%%%%%%%%%%
{\it Introduction.} The quenched approximation is a popular method to estimate QCD quantities on the lattice. By neglecting the fermionic determinant, it is possible to carry out relatively cheap and fast calculations. In the AdS/CFT framework there exists a similar approximation called 'the probe brane approximation', where the flavor-brane backreaction is neglected (the Sakai-Sugimoto model is an example of such a setup \cite{Sakai:2004cn}). Quenching is valid in the 't Hooft limit, where the number of colors is taken to infinity while the number of flavors is kept fixed. In the lattice framework, going beyond the quenched theory
requires powerful computers. In the AdS/CFT framework,  it is difficult to take into account the backreaction of the flavor branes.\footnote{An interesting exception is \cite{Casero:2006pt} where a fully backreacted supersymmetric background was derived.} 

In this short letter we propose a simple method to improve the quenched approximation. The idea is to expand the fermionic determinant in powers of Wilson loops, by using the worldline formalism and to keep the leading and the sub-leading contributions. The outcome is a simple way of calculating flavor-sensitive observables in QCD. A similar idea in lattice gauge theories was suggested a while ago by Sexton and Weingarten \cite{SW}.

We expect the method to be useful especially for holographic calculations. Below we describe our method and demonstrate its usefulness in a simple example.\\

{\it The method.} Consider a calculation of an observable ${\cal O}$ in QCD. In the path integral formalism it can be written, after integration over the fermions, as follows
\beq 
\langle {\cal O} \rangle = \frac{1}{\cal Z} \int DA_\mu \, {\cal O} \, \exp \left( -S_{\rm YM} \right ) \det \left (  i \slsh \!D - m \right ) \, . \label{part1}
\eeq   
The quenched approximation is obtained by omitting the fermionic determinant from the above expression \eqref{part1}
\beq 
\langle {\cal O} \rangle _{\rm YM} = \frac{1}{\cal Z} \int DA_\mu \, {\cal O} \, \exp -S_{\rm YM} \, . \label{part2}
\eeq
Thus $\langle {\cal O} \rangle \approx \langle {\cal O} \rangle _{\rm YM}$ . For certain quantities the above approximation \eqref{part2} turns out to be good. Let us propose a way of improving it.

We use the worldline formalism \cite{Strassler:1992zr} in order to express the fermionic determinant in terms of Wilson loops. The fermionic determinant is related to the Wilson loop as follows
\beq
\det \left (   i\slsh \!D - m \right ) = \exp \Gamma [A] \, ,
\eeq
where
\bea
\label{wlineint}
 \Gamma [A] &=&
-{1\over 2} \int _0 ^\infty {dT \over T}
\nonumber\\[3mm]
 &\times&
\int {\cal D} x {\cal D}\psi
\, \exp
\left\{ -\int _{\epsilon} ^T d\tau \, \left ( {1\over 2} \dot x ^\mu \dot x ^\mu + {1\over
2} \psi ^\mu \dot \psi ^\mu -{1\over 2} m^2 \right )\right\}
\nonumber \\[3mm]
 &\times &  {\rm Tr }\,
{\cal P}\exp \left\{   i\int _0 ^T d\tau
\,  \left (A_\mu \dot x^\mu -\frac{1}{2} \psi ^\mu F_{\mu \nu}  \psi ^\nu
\right ) \right\}  \, ,
\eea
with $x^\mu (0)=x^\mu (T)$.
Thus $\Gamma [A]$ is a sum over (super)-Wilson loops. The sum is over contours of all sizes and shapes. Large contours are, however, suppressed by the quark mass, which serves as an IR cut-off. The sum can be written schematically as $\Gamma [A] = \sum _{\cal C} W$. In this notation the fermionic determinant
is 
\beq
\det \left (   i\slsh \!D - m \right ) = \exp  \sum _ {\cal C} W = \sum _ n \frac{1}{n!} \left ( \sum _ {\cal C} W \right )^n \, . \label{fdet}
\eeq
The quenched theory is obtained by approximating the exponent in \eqref{fdet} by $1$. We propose to improve the approximation by keeping more terms in the above sum \eqref{fdet}. In practise it is easy to add the first term. So, our proposal is
\beq 
 \langle {\cal O} \rangle \approx \langle {\cal O} \rangle _{\rm YM} + \sum _{\cal C} \langle {\cal O } W \rangle _{\rm YM} ^{\rm conn.} \, . \label{proposal}
\eeq
Namely, we suggest to improve the calculation of an observable ${\cal O}$ by adding to the quenched value a sum of its correlator with all possible Wilson loops. The exact weighting is according to \eqref{wlineint}. Another comment is that only connected terms contribute (terms such as  $ \langle {\cal O } \rangle   \langle W \rangle$ are cancelled against the denominator of \eqref{part1}). The present expansion is different than the hopping expansion. In particular, it is valid even when the quark mass is small and, moreover, it can be used in the perturbative regime \cite{Strassler:1992zr}.

In order to obtain a better understanding of the expansion \eqref{fdet} and the proposal \eqref{proposal} let us consider the case of large-$N$ QCD, namely $SU(N_c)$ Yang-Mills with $N_f$ light quarks ($m \ll \Lambda _{\rm QCD}$). The partition functions corresponds to the vacuum energy. We argue that the expansion is Wilson loops is an expansion in powers of $N_f/N_c$. The fully quenched theory contains only gluons, hence the leading large $N_c$ contribution to the vacuum energy is $O(N_c^2)$. In order to 'improve' the calculation, let us use our prescription \eqref{proposal}. A single Wilson loop expectation value contributes as $O(N_c N_f)$. In general, the $n$-th term in the expansion \eqref{fdet} contributes to the vacuum energy as $O\left ( N_c^2 (N_f/N_c)^n \right )$, because each Wilson loop carries a power of $N_f$. Therefore, \eqref{fdet} is a systematic expansion in powers of $N_f/N_c$. Hence the proposal \eqref{proposal} takes into effect the leading $N_f/N_c$ corrections. 

Our idea is very similar to the idea of Sexton and Weingarten \cite{SW} (see also \cite{Duncan:1999xh}) who proposed, in the framework of lattice gauge theories, to expand the determinant in terms of Wilson loops. Note, however, that they focused on improving the action whereas the present method focuses on improving the observable. \\

{\it Example.} Let us consider an example in QCD. We wish to demonstrate the screening effect of the light quarks. The observable ${\cal O}$ that we choose is a circular fundamental Wilson loop of large radius $R$ (with $R \gg \Lambda _{\rm QCD} ^{-1}$).
\begin{figure}[ht]
\centerline{\includegraphics[width=6cm]{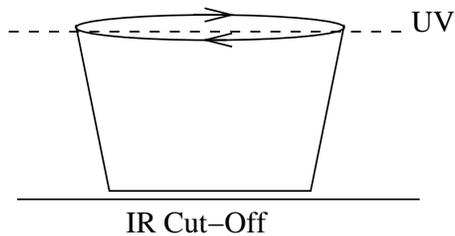}}
\caption{\footnotesize A confining Wilson loop. A holographic calculation by using an AdS with an IR cut-off background.} \label{areafig}
\end{figure}
A calculation of the Wilson loop in the quenched approximation will yield an area law
\beq 
 \langle {\cal O} \rangle _{\rm YM} = \exp -\sigma A \,. \label{area}
\eeq
The above result \eqref{area} can be obtained either by a lattice simulation, or by a lattice strong coupling expansion calculation \cite{Wilson:1974sk}, or by a holographic calculation of the Wilson loop \cite{Maldacena:1998im} in a confining background \cite{Brandhuber:1998er}. Since the theory is quenched string breaking effects of the full theory are not visible.

Let us use our method to improve the result \eqref{area}. We can demonstrate our method by using either the lattice strong coupling expansion or by a holographic calculation. We use the latter. A similar lattice analysis was carried out in \cite{Joos:1983qb}. For simplicity let us choose as a confining Yang-Mills vacuum an AdS metric with an infra-red cut-off
\beq
ds^2 = {du^2 \over u^2} + u^2 dx_i ^2 \,\, ; \,\, u_{\rm min} < u < \infty \, .
\eeq 
 A large Wilson loop in this background will exhibit an area law \eqref{area} since the string worldsheet will immediately drop from the AdS boundary to the IR cut-off $u=u_{\rm min}$ and will 'rest' there, see figure 1.

We now wish to calculate the second term in \eqref{proposal}. The connected two-point function of two Wilson loops, is given by the minimal area worldsheet whose boundaries are the Wilson loops \cite{Berenstein:1998ij},
\beq
 \langle {\cal O } W \rangle _{\rm YM} ^{\rm conn.} = \exp -S_{\rm N.G.} \, . \label{NG}
\eeq
While we formally need to sum over all possible Wilson loop it is clear from the above expression \eqref{NG} that the sum will be dominated by worldsheets with minimal surface. Those are given by Wilson loops which lie close to the circular Wilson loop ${\cal O}$. This is the expected effect of dynamical quarks in the fundamental representation: they create a hole in the QCD-string worldsheet. The minimal area is shown in figure 2 below. 
\begin{figure}[ht]
\centerline{\includegraphics[width=6cm]{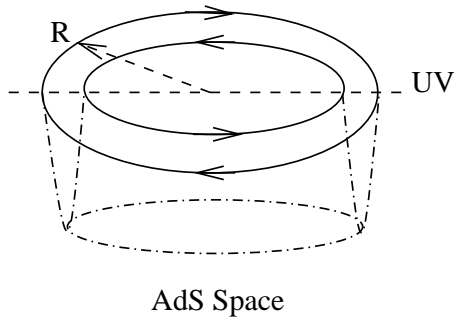}}
\caption{\footnotesize A correlation function of two Wilson loops $\langle W_1 W_2 \rangle$.} \label{WWfig}
\end{figure}
Since the metric near the boundary is AdS, the minimal area is given by
\beq
S_{\rm N.G.} \propto \frac{R}{L} \, ,
\eeq
where $L$ is the difference between the radii of the two Wilson loops. Note that the minimal area is proportional to the perimeter of ${\cal O}$. Thus 
\beq
 \sum _{\cal C} \langle {\cal O } W \rangle _{\rm YM} ^{\rm conn.} \sim \exp -\mu R \, . \label{perimeter}
\eeq
Since the sum is over all sizes of Wilson loops, the value of $\mu$ is expected to be $\mu \sim \Lambda _{\rm QCD}$. Note that we work in the light quark limit. In this limit it is 'easy' to produce pairs and to screen the external charges. When the quark mass is large, the exponent $\exp -m^2 T$ in \eqref{wlineint} suppresses large Wilson loops and it is 'difficult' to create the large Wilson loop that screens ${\cal O}$. In other words, when $m > \Lambda _{\rm QCD}$ it is difficult to produce pairs.
   
Altogether the expectation value of the circular Wilson loop is
\beq
 \langle {\cal O } \rangle = c_1 N_c \exp -\sigma A + c_2 N_f \exp -\mu R \, . \label{result}
\eeq 
$c_1$ and $c_2$ are $O(1)$ coefficients. The prefactor $N_c$ arises since the Wilson loop expectation value is proportional to dim R \cite{Armoni:2006ri}. The correction is proportional to $N_f$ because the expansion is in powers of $N_f/N_c$, as discussed in the previous section. The above expression \eqref{result} is exactly what we expect in QCD. In the absence of quarks, or in the limit $N_c \rightarrow \infty $, fixed $N_f$, the theory exhibits an area law. Incorporating quarks leads to screening. This is reflected by the second term in \eqref{result}. Indeed, when $\Lambda _{\rm QCD}R > \log \frac{N_c}{N_f}$ the second term in \eqref{result} dominates and the QCD-string breaks. 

It will be interesting to carry out a lattice simulation which uses our method, in order to observe the QCD-string breaking. \\

{\it Conclusions.} In this short note we have presented a method to calculate
dynamical quark effects by using the lattice quenched theory (or closed string theory on a curved background). We expect the prescription \eqref{proposal} to be useful for many other calculations as well.

We wish to conclude the paper by proposing an extension of the AdS/CFT dictionary for calculating dynamical matter effects. The large-$N$ (super-)Yang-Mills theory partition function is dual to type II string theory on a curved manifold ${\cal Z}=\exp -S_{\rm II}$ \cite{adscft}. Since Wilson loops are dual to string worldsheets that terminate on the boundary of the AdS (or other) background \cite{Maldacena:1998im}, we propose that the partition function of large-$N$ SU($N$) Yang-Mills theory with an extra massless flavor in the fundamental representation is given by
\beq
{\cal Z}=\exp \tilde \Gamma \, ,
\eeq
where $\tilde \Gamma$ is the partition function of a string whose worldsheets terminate on the boundary of the space manifold
\beq
\tilde \Gamma = \int {\cal D} x {\cal D} g \, \exp \left (-{1\over 4\pi \alpha '}\int d^2 \sigma \sqrt g g^{\alpha \beta} \partial _\alpha x^\mu \partial _\beta x^\nu G_{\mu \nu} \right ) \, . \label{gammatilde} 
\eeq 
To be precise, $\tilde \Gamma$ is a sum over connected worldsheets with $0,1,2,...$ holes on the (AdS, or other) boundary. \\

{\it Acknowledgements.} I wish to thank S. Hands, C. Hoyos, B. Lucini, C. \nunez, A. Patella, and A. Rago  for fruitful discussions. I am supported by the PPARC advanced fellowship award.

%%%%%%%%%%%%%%%%%%%%%%%%%%%%%%%%%%%%%%%%%%%%%%%%%%%%%%%%%%%%%%%%%%%%%%%%%%%%%%%

\end{document}